\documentclass[aps,prb,reprint,longbibliography,floatfix,nofootinbib]{revtex4-2}

\usepackage[T1]{fontenc}
\usepackage[utf8]{inputenc}
\usepackage{amsmath,amssymb,mathtools}
\usepackage{graphicx}
\usepackage{float}
\usepackage{xcolor}
\usepackage{microtype}
\usepackage{silence}
\usepackage[colorlinks=true,linkcolor=blue,citecolor=blue,urlcolor=blue]{hyperref}
\usepackage{orcidlink}

\graphicspath{{figures/}}

\newcommand{\ii}{\mathrm{i}}
\newcommand{\dd}{\mathrm{d}}
\newcommand{\Li}{\operatorname{Li}}

\newcommand{\T}{\mathcal{T}}
\newcommand{\Tconn}{\widetilde{\T}_A}
\newcommand{\K}{\mathcal{K}}

\newcommand{\Qproj}{\mathcal{P}}
\newcommand{\ceff}{c_{\mathrm{eff}}}
\newcommand{\Np}{N_{\mathrm{pockets}}}

\begin{document}

\title{Transfer-matrix functions for algebraically decaying interactions \texorpdfstring{\\}{ } in variational infinite matrix product states}

\author{Qi Yang\,\orcidlink{0000-0002-7325-3100}}
\email{qiyang@mail.ustc.edu.cn}
\affiliation{Institute for Theoretical Physics, University of Amsterdam, 1098 XH Amsterdam, The Netherlands}

\date{\today}

\begin{abstract}
Variational infinite matrix product state (iMPS) calculations usually make Hamiltonians with algebraically decaying interactions compatible with standard MPO algorithms by first replacing the target Hamiltonian with a finite-pole sum-of-exponentials surrogate, thereby introducing a Hamiltonian-representation residual. We formulate the fixed-$D$ variational energy without introducing such a surrogate. For a fixed finite-$D$ MPS, the algebraic tail can be summed directly through the connected transfer matrix: the tail $e^{\ii Qr}/r^\alpha$ is represented by the matrix function $F_{\alpha,Q}(\Tconn)$, with $F_{\alpha,Q}(z)=\Li_\alpha(e^{\ii Q}\,z)/z$. We evaluate the resulting matrix-function action using a Krylov method and obtain stable gradients by combining a Fr\'echet adjoint with implicit fixed-point differentiation.  Benchmarks on long-range free fermions and the inverse-square Heisenberg family, including the Haldane--Shastry point, validate the transfer-matrix-function formulation.  A long-range Ising-chain calculation illustrates a practical consequence of avoiding a finite-pole Hamiltonian representation. At a fixed, independently known critical field, finite-pole surrogate Hamiltonians can bias a critical diagnostic away from criticality, whereas the matrix-function calculation retains the expected critical signatures of the target algebraic Hamiltonian.
\end{abstract}

\maketitle

\section{Introduction}

Algebraically decaying interactions arise in many settings in one-dimensional quantum physics, including inverse-square spin chains \cite{Haldane1988,Shastry1988}, RKKY-type exchange \cite{Ruderman1954,Kasuya1956,Yosida1957,ArguelloLuengo2022}, and long-range interactions in quantum simulators \cite{Defenu2023,Monroe2021,Browaeys2020,Lahaye2009,Micheli2006}.  We use $J(r)=\cos(Qr)/r^\alpha$ as the elementary algebraic tail: $Q=0$ gives the nonoscillatory case, whereas nonzero $Q$ covers staggered and incommensurate exchange.  Finite superpositions of such tails are handled by linearity.

The practical difficulty for matrix product state (MPS) methods \cite{Verstraete2008,Vidal2007,Schollwoeck2011,Orus2014} is that standard matrix product operators (MPOs) naturally represent finite-range couplings and finite sums of exponentials rather than algebraic tails.  The usual workaround therefore replaces the target Hamiltonian with a finite sum of exponentials (SOE) encoded as an MPO \cite{Crosswhite2008,Pirvu2010,Zaletel2015,Hubig2017,LiORourkeChan2019}.  This replacement is not innocuous in the infrared: any finite SOE eventually has an exponential tail and therefore defines a short-range Hamiltonian at asymptotically large distances, even if it approximates the algebraic coupling accurately over a chosen finite window.  Thus the represented Hamiltonian is a finite-pole surrogate, and increasing the MPS bond dimension $D$ improves the variational description of that surrogate rather than removing the representation-level modification of the target algebraic problem.

This distinction is especially important in power-law long-range systems, where the decay exponent itself controls the crossover between long-range and short-range criticality.  Recent high-precision studies of two-dimensional classical models and their $(1+1)$-dimensional quantum counterparts have shown that the crossover boundary, finite-size corrections, and even the interpretation of numerical signatures are delicate \cite{LuijtenBlote2002,BlanchardPiccoRajabpour2013,AngeliniParisiRicciTersenghi2014,HoritaSuwaTodo2017,ShirataniTodo2024,KimKimKim2024NQS,XiaoYaoZhangFanDeng2025XYFerromagnet,YaoXiaoZhangDengFan2025NonclassicalXY}.  In such problems, a finite-pole representation can mix the physical long-range-to-short-range crossover with an artificial crossover built into the Hamiltonian representation.  This motivates a formulation in which the variational state is evaluated with respect to the target algebraic Hamiltonian, without first replacing it by a short-range surrogate.

\begin{figure}[t]
\centering
\includegraphics[width=0.86\columnwidth]{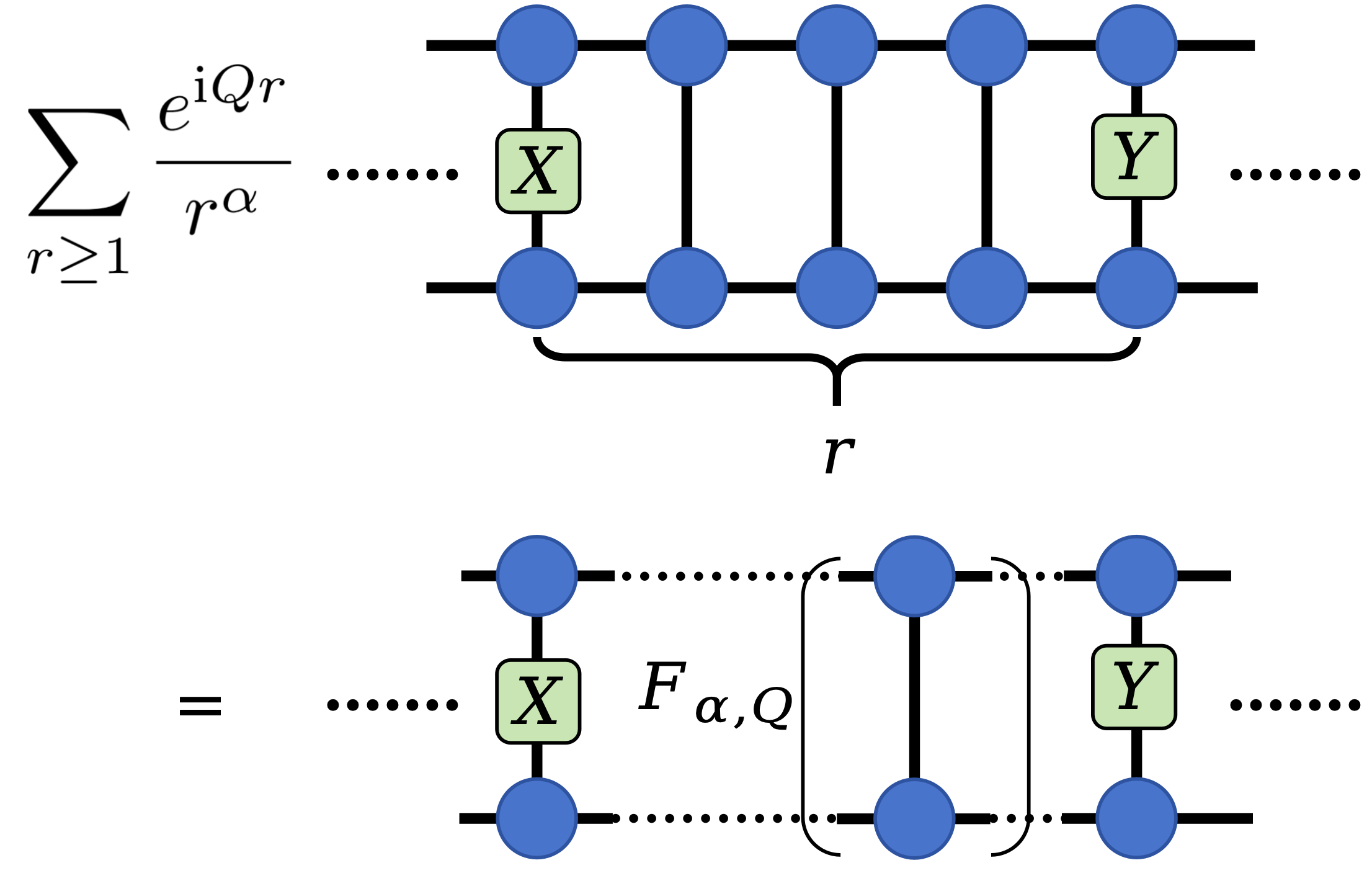}
\caption{Tail summation with transfer matrix function. The connected algebraic tail is evaluated by applying $F_{\alpha,Q}$ [Eq.~\eqref{eq:kernel-intro}] directly to the connected transfer matrix $\Tconn$, rather than by first fitting the tail with a finite-pole MPO representation.}
\label{fig:concept}
\end{figure}

Thus, the objective is not to improve a finite-pole surrogate by adding more poles, but to keep the target Hamiltonian unchanged. For a fixed-$D$ iMPS, connected two-point functions are generated by successive powers of the connected transfer matrix $\Tconn$.  The coefficients of the algebraic interaction can then be summed in the transfer channel, before any real-space truncation or exponential fit is introduced, giving the scalar transfer kernel
\begin{equation}
F_{\alpha,Q}(z)=\sum_{n=0}^{\infty}
  \frac{e^{\ii Q(n+1)}}{(n+1)^\alpha}\,z^n =\frac{\Li_\alpha(e^{\ii Q}\,z)}{z}.
\label{eq:kernel-intro}
\end{equation}
As illustrated in Fig.~\ref{fig:concept}, this kernel is applied to $\Tconn$; here $\Li_\alpha$ is the polylogarithm and $F_{\alpha,Q}(0)=e^{\ii Q}$.

This distinction is central: an SOE-MPO approximation changes the Hamiltonian before the variational problem is solved.  In transfer-matrix language, it replaces the polylogarithmic kernel by a finite rational kernel, with an accuracy controlled by the number of poles and the fitting window.  In the present construction, the kernel is not fitted.  For each source vector $b$, we approximate the finite-dimensional action $F_{\alpha,Q}(\Tconn)b$ with a Krylov method and differentiate that same action through a Fr\'echet adjoint.  Consequently, at fixed $D$, and up to numerical evaluation tolerances, the variational energy is the energy of the target algebraic Hamiltonian: no real-space cutoff, finite-exponential tail fit, or fitted-pole extrapolation enters the Hamiltonian definition.

The remainder of the paper is organized around this separation between the target Hamiltonian and its numerical evaluation.  Sec.~\ref{sec:methods} derives the transfer-matrix-function energy, describes the Krylov action, and explains the variational optimization and Fr\'echet-adjoint gradient.  Sec.~\ref{sec:results} tests the evaluator on long-range free fermions and inverse-square spin chains and then uses a long-range Ising chain to compare the target algebraic Hamiltonian with finite-pole SOE surrogates.  Sec.~\ref{sec:discussion} discusses the scope and limitations of using transfer-matrix-function actions in place of finite-pole Hamiltonian representations.

\section{Methods}
\label{sec:methods}

\subsection{The long-range tail as a transfer-matrix function}
\label{sub:channel}

The formulas below are written for a single-site, translation-invariant, uniform MPS with tensor $A^s$ of bond dimension $D$ and physical dimension $d$; the extension to larger unit cells is straightforward.  The transfer matrix $\T_A$ contracts $A$ with its conjugate $A^\dagger$ over the physical index and acts on the MPS virtual space of bond matrices.  Inserting a one-site operator $O$ defines
\begin{equation}
\T_A^O(W)=\sum_{s,t}O_{st}\,A^tWA^{s\dagger},
\qquad
\T_A\equiv\T_A^{\mathbb I}.
\label{eq:inserted-transfer}
\end{equation}
as a linear map on the $D^2$-dimensional virtual space. Here \(\mathbb I\) denotes the identity matrix.  Let $\langle L|$ and $|R\rangle$ be its left and right dominant fixed points, so that $\langle L|\T_A=\langle L|$, $\T_A|R\rangle=|R\rangle$, and $\langle L|R\rangle=1$.  We assume a unique dominant fixed point that generates the disconnected part of two-point functions.  We remove it with the projector $\Qproj=\mathbb I-|R\rangle\langle L|$ and define the connected transfer matrix $\Tconn=\Qproj\T_A\Qproj$.  We use the commutation relation $\left[\T_A,\Qproj\right]=0$ below.

For $O=X,Y$, define the one-site expectation value and centered operator as $\langle O\rangle=\langle L|\T_A^O|R\rangle$ and $\widetilde O=O-\langle O\rangle\mathbb I$.  The connected readout and source are
\begin{equation}
\langle L_X| =\langle L|\T_A^{\widetilde X}\Qproj,\quad |R_Y\rangle=\Qproj\T_A^{\widetilde Y}|R\rangle .
\label{eq:readout-source}
\end{equation}

For $r\ge1$, the connected two-point function associated with the centered
insertions is
\begin{equation}
C_{\widetilde X\widetilde Y}(r)
:=\langle \widetilde X_i\widetilde Y_{i+r}\rangle
=\langle L_X|\Tconn^{\,r-1}|R_Y\rangle .
\end{equation}
Hence
\begin{equation}
\langle X_iY_{i+r}\rangle
=\langle X\rangle\langle Y\rangle+C_{\widetilde X\widetilde Y}(r).
\end{equation}
With $F_{\alpha,Q}(M):=\sum_{r\ge1}e^{\ii Qr}M^{r-1}/r^\alpha$, the algebraic
generating sum becomes
\begin{align}
\sum_{r\ge1}\frac{e^{\ii Qr}}{r^\alpha}
\langle X_iY_{i+r}\rangle
&=\Li_\alpha(e^{\ii Q})\langle X\rangle\langle Y\rangle \notag\\
&+\langle L_X|F_{\alpha,Q}(\Tconn)|R_Y\rangle .
\label{eq:connected-tail}
\end{align}
Thus the original tail sum decomposes into disconnected terms
and a connected-channel matrix function $F_{\alpha,Q}(\Tconn)$.

By linearity, it is enough to write the energy density for one real cosine interaction term, $J\cos(Qr)X_iY_{i+r}/r^\alpha$:
\begin{align}
&E_{\alpha,Q}[X,Y]
\equiv \sum_{r\ge1}\frac{J\cos(Qr)}{r^\alpha}\langle X_iY_{i+r}\rangle
\label{eq:matrix-function}
\\
&=J\,\operatorname{Re}\!\Big[
  \Li_\alpha(e^{\ii Q})\,\langle X\rangle\langle Y\rangle +\langle L_X|F_{\alpha,Q}(\Tconn)|R_Y\rangle\Big] .\notag
\end{align}
For the disconnected scalar term at $Q=0$, Eq.~\eqref{eq:matrix-function} assumes $\alpha>1$; otherwise it requires centering or another regularization.
Since Eqs.~\eqref{eq:connected-tail} and~\eqref{eq:matrix-function} are identities, the tail contribution is represented exactly at fixed $D$, and the energy carries no tail-representation error; the reformulation removes this source of error entirely, while all other fixed-$D$ errors are inherited from the underlying calculation and thus remain unchanged by the reformulation.

Viewed as a function of the transfer matrix, the discrete tail sum is a $z$-transform, analogous to a Laplace transform in continuous matrix product states \cite{HaegemanThesis2011,HaegemanCMPS2013}.  This viewpoint also connects to SOE-MPO constructions: a single exponential weight $\lambda^r$ gives the rational geometric kernel $\lambda/(1-\lambda z)$, so a finite sum of exponentials gives a finite rational kernel representable by a finite-bond MPO \cite{Crosswhite2008,Pirvu2010,Zaletel2015,Hubig2017,LiORourkeChan2019}.  The algebraic tail instead gives the non-rational kernel $F_{\alpha,Q}(z)=\Li_\alpha(e^{\ii Q}\,z)/z$, which we apply directly rather than replacing it by a rational fit.

\subsection{Krylov evaluation of the matrix-function action}
\label{sub:krylov}
We focus on the connected matrix-function contribution in Eq.~\eqref{eq:matrix-function},
$\langle L_X|F_{\alpha,Q}(\Tconn)|R_Y\rangle$.  A formal route would be to diagonalize the transfer matrix $\Tconn$.  If
$\Tconn=\sum_{i=1}^{D^2}\lambda_i |x_i\rangle \langle \tilde x_i|$, with right eigenvectors $|x_i\rangle$ and dual left eigenvectors $\langle \tilde x_i|$, then
\begin{equation}
\begin{aligned}
\langle L_X|F_{\alpha,Q}(\Tconn)|R_Y\rangle
&=\sum_{i=1}^{D^2}F_{\alpha,Q}(\lambda_i)\,
  \langle L_X|x_i\rangle
  \langle \tilde x_i|R_Y\rangle .
\end{aligned}
\label{eq:dense-spectral-readout}
\end{equation}
This expression makes the structure transparent: the long-range tail is a spectral readout of scalar kernel values, but the identity is more useful as a conceptual guide than as an algorithm.  The full channel has dimension $D^2$, so storing it densely costs $O(D^4)$ and diagonalizing it costs $O(D^6)$.  More importantly, $\Tconn$ is generally non-normal---it is the same finite-$D$ transfer matrix that sets iMPS correlation lengths and finite-entanglement scaling \cite{Zauner2015,Stojevic2015,Vanhecke2019}---so the spectral weights in Eq.~\eqref{eq:dense-spectral-readout} can be badly conditioned when eigenvectors are nearly linearly dependent \cite{Higham2008,TrefethenEmbree2005}.

We therefore keep the spectral readout but move it to the Krylov projected subspace.  Let $V_m$ be the orthonormal Arnoldi basis generated from $|R_Y\rangle$, so that $\K_m(\Tconn,|R_Y\rangle)=\operatorname{span}\{\Tconn^{\,j}|R_Y\rangle\}_{0\le j<m}$, and let $H_m=V_m^\dagger\Tconn V_m$ be the upper-Hessenberg compression.  If $(\theta_i,|u_i\rangle,\langle\tilde u_i|)$ are biorthogonal eigenpairs of this $m\times m$ matrix, the resulting readout is the projected spectral sum
\begin{equation}
\begin{aligned}
\langle L_X|F_{\alpha,Q}(\Tconn)|R_Y\rangle
&\approx
\sum_{i=1}^{m}F_{\alpha,Q}(\theta_i)\,
  \langle L_X|V_m|u_i\rangle\\
&\qquad\times
  \langle\tilde u_i|V_m^\dagger|R_Y\rangle .
\end{aligned}
\label{eq:krylov-ritz-readout}
\end{equation}
This is the Krylov-space analogue of Eq.~\eqref{eq:dense-spectral-readout}: the full eigenvalues $\lambda_i$ of $\Tconn$ are replaced by the Ritz values $\theta_i$ of $H_m$, and the weights are projected through $V_m$.  The full transfer matrix enters only through the $m$ Arnoldi matrix--vector products used to build $V_m$ and $H_m$; after projection, the algebraic tail enters through scalar evaluations of $F_{\alpha,Q}$ on the Ritz values, as detailed in Appendix~\ref{app:scalar-kernel} \cite{Saad1992,Guettel2013,Higham2008}.

In terms of computational complexity, each Arnoldi iteration applies $\Tconn$ once at $O(D^3)$, so $m$ steps cost $O(mD^3)$, compared with the $O(\chi_{\mathrm{MPO}}D^3)$ contraction cost of an SOE-MPO baseline of bond $\chi_{\mathrm{MPO}}$.  The Krylov dimension $m$ controls the numerical evaluation of the matrix-function action; it is not a real-space cutoff or pole count.

\begin{figure}[t]
\centering
\includegraphics[width=\columnwidth]{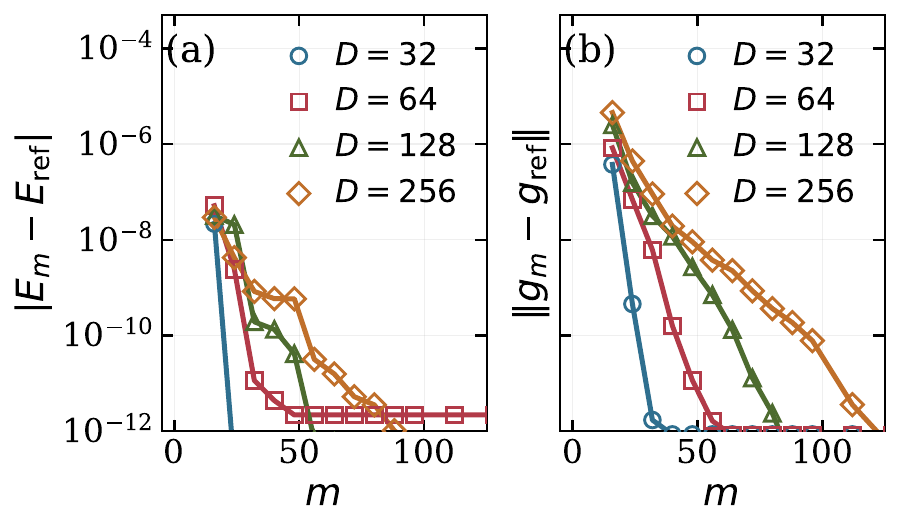}
\caption{Accuracy versus Krylov dimension $m$ for converged finite-$D$ iMPS of the nonoscillatory infinite Haldane--Shastry model, with $D=32,64,128,256$. Reference values $E_{\rm ref}$ and $g_{\rm ref}$ are fixed-MPS large-$m$ references from the converged $m_{\rm ref}=512$ calculation; $E_{\rm ref}$ is independently certified by direct real-space summation (Sec.~\ref{sec:methods}). (a) Energy error.  (b) Projected-gradient error norm.}
\label{fig:hs-krylov}
\end{figure}

The remaining question is how large the Krylov subspace dimension $m$ must be for the MPS states studied in practice.  Fig.~\ref{fig:hs-krylov} isolates that question by freezing optimized critical finite-$D$ iMPS states for the infinite Haldane--Shastry model and reevaluating the same fixed MPS states as $m$ is varied.  The reference values $E_{\rm ref}$ and $g_{\rm ref}$ are large-$m$ energy and projected-gradient references, taken from the converged $m_{\rm ref}=512$ calculation.  For the energy we also check the reference independently: direct real-space summation of $\tfrac14\sum_{r\ge1}r^{-2}\langle\vec\sigma_0\!\cdot\!\vec\sigma_r\rangle$ through ordinary transfer-matrix correlators---with no matrix-function action---reproduces $E_{\rm ref}$. The resulting curves show that moderate Krylov dimensions reduce both the energy and projected-gradient errors to the plotted numerical floor over the tested $D=32$--$256$ window.

\subsection{Variational optimization}
\label{sub:opt}

We minimize the energy variationally, using gradients obtained by reverse-mode automatic differentiation~\cite{Liao2019,Xie2020,Francuz2025}. The implicit fixed-point part follows the dominant-eigensolver pullback of Ref.~\cite{Xie2020}. The key technical additions are two differentiable primitives introduced in this work, extending previous approaches. First, we differentiate the matrix-function action $y=F(M)b$ through its Fr\'echet derivative rather than through the Krylov recurrence \cite{Kraemer2024Gradients}.  For an output cotangent $\bar{y}$, the source cotangent is $\bar{b}=F(M)^\dagger\bar{y}$; the channel cotangent is the adjoint Fr\'echet action, evaluated with independent right and left Krylov spaces of the kernel (Appendix~\ref{app:frechet-adjoint}). Second, we use the Daleckii--Krein formula to avoid Lorentz broadening when Ritz values of $H_m$ become degenerate (Appendix~\ref{app:daleckii}).

The resulting raw gradient is projected to the uniform MPS tangent space \cite{Vanderstraeten2019TangentSpace}, and the energy is minimized using Riemannian L-BFGS optimization on the manifold of isometric tensors \cite{Hauru2021}, implemented with TensorKit, MPSKit, OptimKit, and TensorKitManifolds \cite{TensorKit,MPSKit,OptimKit,TensorKitManifolds}.  For the largest-$\xi_D$ optimizations, we damp the Riemannian preconditioner used in the MPSKit optimization stack, interpolating it toward the identity before the L-BFGS update; without this damping, the preconditioned gradient can become anomalously large even when the raw projected gradient is well controlled.

\section{Results}
\label{sec:results}

Unless stated otherwise, we set $\alpha=2$.  Finite-$D$ data are plotted against the transfer correlation length
\begin{equation}
\xi_D=-\frac{1}{\ln|\lambda_2|},
\label{eq:transfer-xi}
\end{equation}
where $\lambda_2$ is the leading subdominant eigenvalue of the relevant transfer channel: the connected channel for ordinary correlators, and the string or charged channel for Jordan--Wigner strings.

\subsection{Long-range free fermions with Jordan--Wigner strings}
\label{sub:fermions}

\begin{figure}[t]
\centering
\includegraphics[width=\columnwidth]{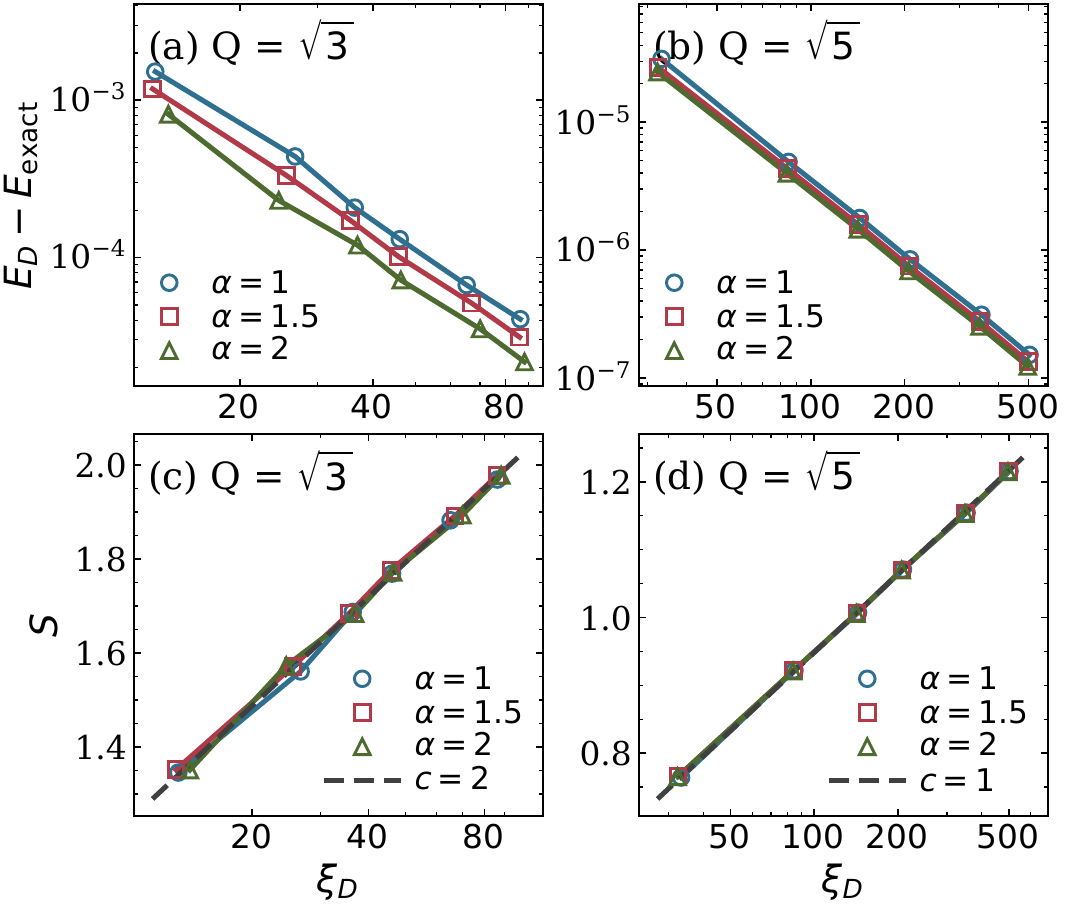}
\caption{Long-range free-fermion string-channel benchmark.  All panels use a logarithmic horizontal axis for $\xi$ with $D$ ranging from $16$ to $128$.  (a,b) $E(D_\xi)-E_{\mathrm{exact}}$ relative to the exact thermodynamic reference.  (c,d) Entropy scaling for $Q=\sqrt3$ and $Q=\sqrt5$; dashed lines are visual guides.}
\label{fig:jw}
\end{figure}

We first consider the long-range spinless-fermion Hamiltonian
\begin{equation}
\begin{aligned}
H_c
&=\sum_i\sum_{r\ge1}
  \frac{\cos(Qr)}{r^\alpha}
  \left(c_i^\dagger c_{i+r}+c_{i+r}^\dagger c_i\right).
\end{aligned}
\label{eq:fermion-hamiltonian}
\end{equation}
After a Jordan--Wigner transformation, the hopping endpoints are joined by a parity string \cite{JordanWigner1928,LiebSchultzMattis1961}.  With $\hat n=(1-\sigma^z)/2$, $\Pi=(-1)^{\hat n}=\sigma^z$, and $c_j=(\prod_{m<j}\Pi_m)\sigma_j^+$, for $r>0$,
\begin{equation}
c_i^\dagger c_{i+r}
=\sigma_i^-
\left(\prod_{m=i+1}^{i+r-1}\Pi_m\right)
\sigma_{i+r}^+.
\label{eq:jw-string}
\end{equation}
At the transfer-matrix level, the string replaces the ordinary propagator by $\T_A^\Pi$, so that
\begin{equation}
\langle c_i^\dagger c_{i+r}\rangle
=\langle L|\T_A^{\sigma^-}
(\T_A^\Pi)^{r-1}
\T_A^{\sigma^+}|R\rangle.
\label{eq:jw-transfer}
\end{equation}
The tail is therefore summed by applying $F_{\alpha,Q}$ to the string transfer channel $\T_A^\Pi$ (or the corresponding charged channel in the symmetry-native implementation).  Because the endpoints are charged, the ordinary neutral disconnected piece vanishes by symmetry rather than by a mean-field subtraction.

This quadratic model has the exact thermodynamic dispersion
\begin{equation}
\begin{aligned}
\varepsilon(k)
&=2\sum_{r\ge1}\frac{\cos(Qr)\cos(kr)}{r^\alpha}\\
&=\operatorname{Re}\!\Big[
  \Li_\alpha(e^{\ii(Q+k)})
 +\Li_\alpha(e^{\ii(Q-k)})\Big].
\end{aligned}
\label{eq:jw-dispersion}
\end{equation}
At zero chemical potential, the thermodynamic ground state fills the modes with $\varepsilon(k)<0$, giving the reference energy used here; at fixed density, one would instead fill modes up to the density-determined Fermi level.  Here $\Np$ denotes the number of occupied Fermi intervals.  Fig.~\ref{fig:jw} compares the variational energies with this reference for $D\le128$; the energy panels show convergence toward the exact thermodynamic values over the plotted $D$ window.

The entanglement entropy scaling checks $\ceff\simeq\Np$: over the plotted $D\le128$ window, the fitted values are $\ceff=2.00(5),1.97(2),1.99(5)$ for $Q=\sqrt3$ at $\alpha=1,1.5,2$, and $\ceff=0.995(4),0.985(4),0.986(4)$ for $Q=\sqrt5$, with parentheses denoting one-standard-error fit uncertainty in the last shown digits.  This is consistent with the expectation that each disconnected Fermi interval contributes one gapless pocket \cite{JinKorepin2004,CalabreseMintchevVicari2011}.

\subsection{Inverse-square Heisenberg family}
\label{sub:hs}

We next consider the spin-$1/2$ inverse-square Heisenberg family
\begin{equation}
H_{\mathrm H}(Q)
=\frac14\sum_i\sum_{r\ge1}
  \frac{\cos(Qr)}{r^2}\,
  \vec\sigma_i\!\cdot\!\vec\sigma_{i+r}.
\label{eq:heis}
\end{equation}
The nonoscillatory point $Q=0$ is the thermodynamic Haldane--Shastry (HS) chain \cite{Haldane1988,Shastry1988,HaldaneHa1992,CiracSierra2010}, with exact energy density $E_{\mathrm{HS}}=-\pi^2/24$, known spin correlators \cite{GebhardVollhardt1987,TuNielsenCiracSierra2014,TuNielsenSierra2014}, and $SU(2)_1$ low-energy theory with central charge $c=1$ \cite{HaldaneHa1992,CiracSierra2010}.  This makes $Q=0$ a stringent exact benchmark for the matrix-function evaluator.

For $Q\neq0$, the oscillatory factor introduces an additional length scale $2\pi/Q$ into the inverse-square exchange and breaks the special structure of the Haldane--Shastry point.  To our knowledge, this finite-$Q$ family has not been characterized by an exact solution.  It is therefore unclear a priori whether the accessible finite-$D$ states should show the same conformal finite-entanglement structure as the $Q=0$ chain, or whether the additional scale eventually drives the system away from a CFT description.  We do not attempt a full phase characterization of this family here.  Instead, the finite-$Q$ points provide nontrivial tests of the complex kernel $F_{\alpha,Q}$ and finite-entanglement scaling away from the exactly solvable point.

For the finite-$Q$ spin-chain optimizations we use a Marshall-folded local frame \cite{Marshall1955}.  This is only a choice of variational coordinates; it does not modify the Hamiltonian or the matrix-function evaluator.  In practice, it prevents the one-site optimization from converging to cat-like or effectively doubled representations of the same branch, thereby saving the bond dimension that would otherwise be spent on this avoidable doubling.  We do not interpret the folding as a physical assumption about an exact Marshall sign rule at finite $Q$.

\begin{figure}[t]
\centering
\includegraphics[width=\columnwidth]{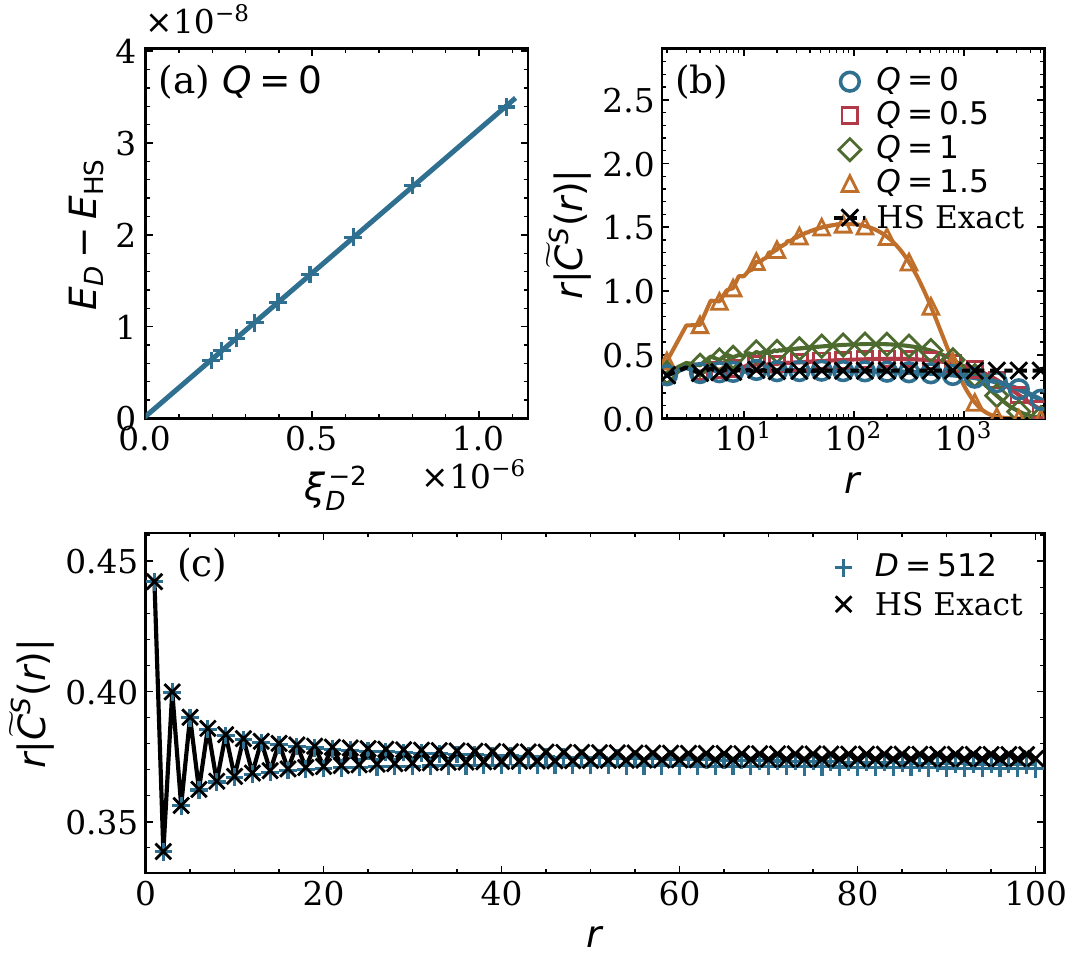}
\caption{Inverse-square Heisenberg energy and correlator benchmarks.  (a) $Q=0$ Haldane--Shastry energy residual versus $\xi_D^{-2}$; the fit uses $D=256$--$512$.  (b) Normalized connected spin-vector correlators $\widetilde C^S_Q(r)=\langle\vec S_0\!\cdot\!\vec S_r\rangle_c$ at $D=512$ for $Q=0,0.5,1,1.5$; the dashed guide is the HS exact reference.  (c) Direct $Q=0$ pointwise comparison of $r|\widetilde C^S(r)|$ against the exact spin-vector HS result $|\widetilde C^S_{\mathrm{HS}}(r)|=3\,\operatorname{Si}(\pi r)/(4\pi r)$.}
\label{fig:hs-precision}
\end{figure}

Fig.~\ref{fig:hs-precision}(a) checks the $Q=0$ energy against the HS reference.  A linear fit in $\xi_D^{-2}$ over $D=256$--$512$ gives $E_\infty-E_{\mathrm{HS}}=+2.2\times10^{-10}$, showing that the transfer-matrix-function evaluator reaches the exact thermodynamic energy with very high precision.  Panels (b,c) provide a complementary correlator check.  At $Q=0$, the pointwise connected spin-vector correlator agrees with the Gebhard--Vollhardt thermodynamic result $|\widetilde C^S_{\mathrm{HS}}(r)|=|\langle\vec S_0\!\cdot\!\vec S_r\rangle|=3\operatorname{Si}(\pi r)/(4\pi r)$ \cite{GebhardVollhardt1987}; equivalently the Pauli-dot magnitude is $|\langle\vec\sigma_0\!\cdot\!\vec\sigma_r\rangle|=3\operatorname{Si}(\pi r)/(\pi r)$.  The finite-$Q$ curves in panel (b) illustrate that the oscillatory inverse-square chains are not simply reproducing the HS correlator: already at short and intermediate distances their connected correlators deviate visibly from the $Q=0$ exact form.

\begin{figure}[t]
\centering
\includegraphics[width=\columnwidth]{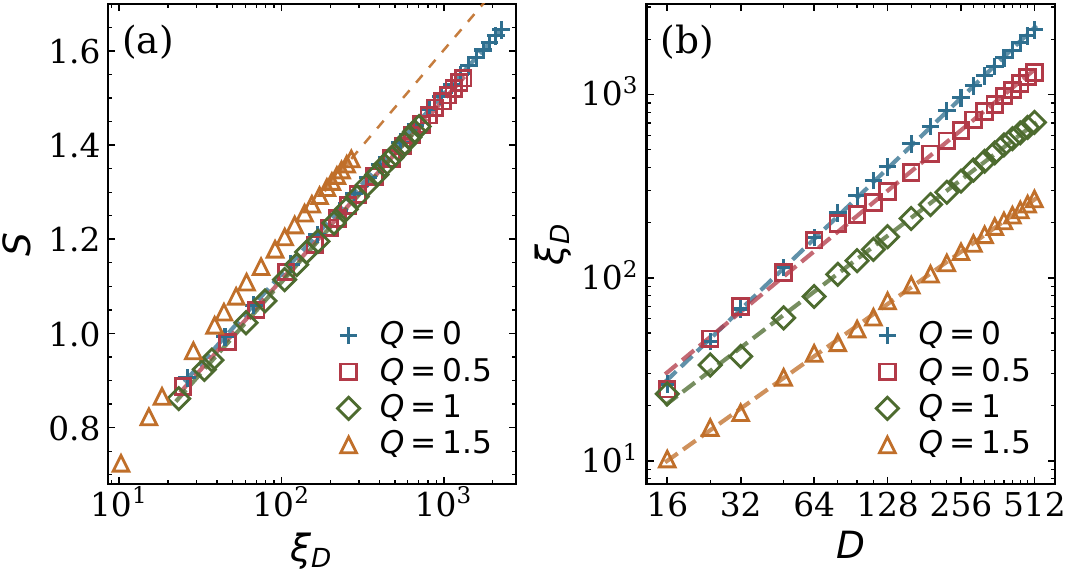}
\caption{Finite-entanglement scaling for Eq.~\eqref{eq:heis} at $\alpha=2$. (a) Entanglement entropy $S$ versus $\xi_D$ for $Q=0,0.5,1,1.5$, with log-linear fits capped at $D\le512$.  (b) Correlation length $\xi_D$ versus bond dimension $D$ on logarithmic axes, using the same $D\le512$ window.}
\label{fig:escorr}
\end{figure}

The finite-entanglement scaling in Fig.~\ref{fig:escorr}(a) gives another view of the same finite-$Q$ states.  For a finite-$D$ iMPS approximation to a critical state, the entanglement entropies are fitted with the finite-entanglement form~\cite{CalabreseCardy2004,Tagliacozzo2008,Pollmann2009,Pirvu2012,Stojevic2015} $S_D=s_0+(c/6)\ln\xi_D+\cdots$.  The entropy slopes remain close to the $c=1$ line throughout the accessible window. The fitted values are $\ceff=0.998(1),0.997(3),1.014(2),1.055(7)$ for $Q=0,0.5,1,1.5$; the $Q=1.5$ fit uses only the largest six available $D$ values.

Fig.~\ref{fig:escorr}(b) shows that the corresponding $\xi_D$-versus-$D$ curves grow cleanly rather than saturate, consistent with gapless behavior over the accessible $D$ range. Together with the correlator deformation in Fig.~\ref{fig:hs-precision}(b), these data suggest that the finite-$Q$ inverse-square chains may remain gapless while no longer being described by the unmodified HS correlator structure. Establishing the continuum theory, possible crossover scales, and the fate of conformal invariance at finite $Q$ would require a more systematic study. Such a study is beyond the scope of the present benchmark.

\subsection{Long-range ferromagnetic Ising}
\label{sub:lrising}

For the SOE comparison, we consider the long-range transverse-field ferromagnetic Ising chain
\begin{equation}
\begin{aligned}
H
&=-\sum_i\sum_{r\ge1}
  \frac{\sigma_i^x\sigma_{i+r}^x}{r^\alpha}
  -\Gamma\sum_i\sigma_i^z .
\end{aligned}
\label{eq:lrising}
\end{equation}

The critical field $\Gamma_c$ and exponents vary continuously with $\alpha$ \cite{Defenu2023,Sak1973,Koziol2021,Defenu2015}.  At $\alpha=1.5$ the chain lies in the long-range mean-field regime \cite{FisherMaNickel1972,DuttaBhattacharjee2001,Defenu2015}.  Writing $\sigma_{\rm LR}=\alpha-1$, the equal-time order correlator at criticality is expected to decay as $C(r)=\langle\sigma_0^x\sigma_r^x\rangle_c\sim r^{-\eta}$ with $\eta=1-\sigma_{\rm LR}/2=0.75$ \cite{DuttaBhattacharjee2001,Defenu2023}.  Thus a compensated correlator $r^{\eta}|C(r)|$ should plateau over distances below the finite-$D$ correlation length.

We evaluate the chain at the high-precision quantum Monte Carlo (QMC) estimate of the critical field, $\Gamma=4.75999$ \cite{Koziol2021}, and contrast the matrix-function calculation with an SOE-MPO baseline.

\begin{figure}[t]
\centering
\includegraphics[width=\columnwidth]{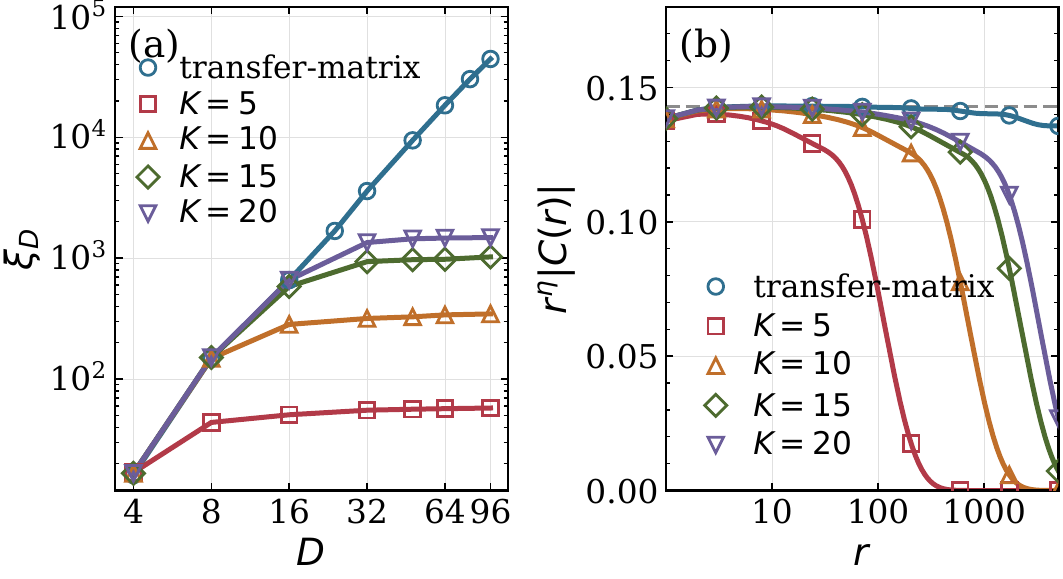}
\caption{Comparison between the SOE-MPO and transfer-matrix-function approaches for Eq.~\eqref{eq:lrising} at $\alpha=1.5$ and the critical field $\Gamma=4.75999$ \cite{Koziol2021}, in the $\langle\sigma^x\rangle$ channel.  (a) Transfer correlation length $\xi_D$ versus $D$ for the converged matrix-function ladder ($D=4$--$96$) and SOE-MPO Hamiltonians with $K=5,10,15,20$ poles fitted to $R_{\max}=10^4$. (b) Compensated connected order correlator $r^{\eta}|C(r)|$ with $\eta=0.75$.  Panel (b) measures the $\sigma_x$ correlator of the $D=96$ matrix-function optimized state and SOE-VUMPS optimized states for $K=5,10,15,20$, respectively; the dashed line is the mean matrix-function value over $10\le r\le100$.}
\label{fig:convergence}
\end{figure}

The baseline fits the tail with a finite sum of exponentials,
\begin{equation}
\frac{1}{r^{\alpha}}
\simeq \sum_{k=1}^{K}c_k\,\lambda_k^{\,r},
\label{eq:soe}
\end{equation}
obtained using the Tensor Network Python (TeNPy) fitter \cite{TeNPy2018} over $1\le r\le R_{\max}$ and encoded as $K$ poles in an MPO. Fig.~\ref{fig:convergence} uses $R_{\max}=10^4$ and $K=5,10,15,20$. Across this sequence the absolute tail residual decreases from $4.4\times 10^{-4}$ to $2.8\times 10^{-6}$. This gives a systematic finite-pole sequence, not a statement about the asymptotic capability of SOE-MPOs: increasing $K$ can systematically reduce the residual and extend the reliable range.

The comparison is intentionally made at the fixed critical field.  A finite-$K$ SOE fit defines a nearby approximate Hamiltonian whose apparent critical field $\Gamma_c(K)$ may depend on $K$. Retuning each SOE-MPO to its own $\Gamma_c(K)$ would test the critical scaling of the corresponding finite-$K$ approximate Hamiltonian, not the original algebraic Hamiltonian at the QMC field. Moreover, extrapolating $\Gamma_c(K)$ with $K$ would require an additional convergence assumption tied to the chosen fit norm and window. Here the question is sharper: does the finite-pole representation preserve the critical observable of the original Hamiltonian at a fixed, independently known field?

Fig.~\ref{fig:convergence} shows that, in the matrix-function calculation at the QMC field, the correlation length $\xi_D$ does not saturate over the plotted $D=4$--$96$ ladder. The compensated connected order-parameter correlator, computed from the $D=96$ state, remains close to its mid-range plateau up to the largest plotted distances. By contrast, for the SOE-approximated Hamiltonians, VUMPS ground states obtained at the same QMC field show finite correlation lengths on the accessible $D$ ladder, far below those obtained from the matrix-function evaluation.  The order-parameter correlator then follows the compensated power-law guide only up to a $K$-dependent range before crossing over below the matrix-function plateau.

Increasing $K$ moves the SOE curves toward the matrix-function curve, as expected for a systematically improved surrogate. The point of the benchmark is that the convergence variable $K$ belongs to the Hamiltonian representation, not to the MPS variational ansatz. In this observable and parameter window, that extra representation layer appears as a bias in the critical correlator rather than as a small uniform energy offset.

\section{Discussion}
\label{sec:discussion}

The central message of this work is that, for a fixed finite-$D$ uniform MPS, an algebraic interaction tail need not be represented by an auxiliary finite-pole Hamiltonian.  After separating the disconnected fixed-point contribution, the remaining connected tail is a matrix function of the connected transfer channel.  The Krylov approximation then controls the numerical evaluation of this finite-dimensional matrix-function action, rather than defining a sequence of surrogate Hamiltonians.  In this formulation, the finite-entanglement approximation and the Hamiltonian-representation approximation are disentangled.

The methodology is not tied to the particular zero-temperature infinite-chain setting used in these benchmarks.  Its two main algorithmic ingredients—Krylov evaluation of transfer-matrix functions and the Daleckii–Krein divided-difference pullback—are general tensor-network primitives.  They can be applied whenever an analytic scalar kernel acts on a transfer matrix, an environment map, or a projected effective channel.  This suggests direct extensions to finite-temperature tensor networks, finite rings with periodic boundary conditions, quasi-one-dimensional geometries, and broader tensor-network algorithms where long-range kernels or nonlocal response functions are naturally expressed as functions of transfer operators.

The remaining limitations are then those of the variational representation and the numerical matrix-function evaluation, rather than of a fitted Hamiltonian tail.  Critical states remain more expensive as the finite-$D$ transfer gap closes. For commensurate ordered states, larger unit cells may still be preferable to avoid cat-like representations that waste bond dimension.  Within these constraints, control of the approximation is shifted back to the tensor-network ansatz and the evaluator tolerance, instead of being built into a finite-pole representation of the Hamiltonian.

\section*{Data Availability}

The data, scripts, reproduction guide, and parameter sets needed to reproduce the figures are available in the project repository~\cite{TransferMatrixIMPSRepository}.

\begin{acknowledgments}
I thank Philippe Corboz for numerous discussions and continued support, Youjin Deng for originally proposing this project and emphasizing its significance, and Honghao Tu, Jutho Haegeman, Kareljan Schoutens, Jean-Sébastien Caux, and Anton Romen for helpful comments and discussions.  This project has received funding from the European Research Council (ERC) under the European Union's Horizon 2020 research and innovation programme (grant agreement No. 101001604).
\end{acknowledgments}

\appendix

\section{Scalar Function Evaluation}
\label{app:scalar-kernel}

The Krylov projection requires evaluating the scalar function $F_{\alpha,Q}(z)=\Li_\alpha(e^{\ii Q}\,z)/z$ and its derivative at Ritz values obtained from small projected matrices.  The branch choices and continuation formulas follow the DLMF polylogarithm entry \cite{DLMFPolylog}. The removable value at the origin is $F_{\alpha,Q}(0)=e^{\ii Q}$.  For $\alpha=1$, we use $\Li_1(z)=-\ln(1-z)$.  For $\alpha=2$, the dilogarithm is evaluated using the Bernoulli series in $u=-\ln(1-z)$ after standard inversion and reflection reductions.  For real noninteger $\alpha>1$, we use the ordinary power series away from the branch point and the continuation
\begin{equation}
\begin{aligned}
\Li_s(e^\mu)
&=\Gamma(1-s)(-\mu)^{s-1}
  +\sum_{k=0}^{\infty}\frac{\zeta(s-k)}{k!}\mu^k .
\end{aligned}
\label{eq:polylog-continuation}
\end{equation}
near the branch point. We use this form for $|\mu|<2\pi$ on the chosen logarithm branch, with the regularized integer limit taken when needed.  The derivative used in the Fr\'echet adjoint is
\begin{equation}
\begin{aligned}
F'_{\alpha,Q}(z)
&=\frac{\Li_{\alpha-1}(e^{\ii Q}\,z)
        -\Li_{\alpha}(e^{\ii Q}\,z)}{z^2}.
\end{aligned}
\label{eq:polylog-derivative}
\end{equation}
again using the regular limit at the origin.

\section{Fr\'echet Adjoint}
\label{app:frechet-adjoint}

The primitive to be differentiated is
\begin{equation}
y=F(M)b ,
\label{eq:app-matfun-primitive}
\end{equation}
where $M$ is a connected transfer matrix or string channel and $b$ is the corresponding source.  Its first variation is
\begin{equation}
\begin{aligned}
\delta y&=F(M)\,\delta b+\mathcal{L}_F(M;\delta M)b,\\
\mathcal{L}_F(M;\delta M)&=
\left.\frac{\dd}{\dd\epsilon}F(M+\epsilon \delta M)\right|_{\epsilon=0}.
\end{aligned}
\label{eq:frechet-variation}
\end{equation}
Equivalently,
\begin{equation}
F\!\left[
\begin{pmatrix}
M&\delta M\\
0&M
\end{pmatrix}
\right]
=
 \begin{pmatrix}
F(M)&\mathcal{L}_F(M;\delta M)\\
0&F(M)
\end{pmatrix},
\label{eq:frechet-block}
\end{equation}
defines the derivative without diagonalizing $M$ \cite{Higham2008,KandolfRelton2017}.

For an incoming cotangent $\bar{y}$, the pullback is defined through the Hilbert--Schmidt pairing; for real-valued energy functionals, a real part is understood:
\begin{equation}
\begin{aligned}
\delta\Phi
&=\langle \bar{y},F(M)\delta b\rangle
  +\langle \bar{y},\mathcal{L}_F(M;\delta M)b\rangle\\
&=\langle \bar{b},\delta b\rangle+\langle \bar{M},\delta M\rangle .
\end{aligned}
\label{eq:frechet-pullback-def}
\end{equation}
Thus
\begin{equation}
\begin{aligned}
\bar{b}&=F(M)^\dagger \bar{y},\\
\langle \bar{M},\delta M\rangle
&=\langle \bar{y},\mathcal{L}_F(M;\delta M)b\rangle
\quad \text{for all } \delta M .
\end{aligned}
\label{eq:frechet-pullback}
\end{equation}
For the polylogarithmic kernel with real $\alpha$, the source action is evaluated as $F_{\alpha,-Q}(M^\dagger)\bar{y}$ on the conjugate branch.  The channel cotangent $\bar{M}$ is computed from the Daleckii--Krein representation of $\mathcal{L}_F$ given in Appendix~\ref{app:daleckii}.

\par\vspace{0.75\baselineskip}
\section{Daleckii--Krein Divided Differences}
\label{app:daleckii}

In an eigenbasis of $M$, the Fr\'echet derivative
$\mathcal{L}_F(M;\delta M)$ is obtained by multiplying the
matrix elements of $\delta M$ entrywise by the divided
differences of the scalar kernel. This is the
Daleckii--Krein formula
\cite{DaleckiiKrein1965,Higham2008}, and it is a core
primitive for reverse-mode differentiation of spectral matrix
functions in tensor-network implementations
\cite{Kraemer2024Gradients,Liao2019}.

We evaluate the corresponding adjoint action in independent
right and left Krylov spaces,
$\K_m^R=\K_m(M,b)$ and $\K_n^L=\K_n(M^\dagger,\bar{y})$.
Here $X^\dagger$ denotes the Hilbert-space adjoint of a linear
operator, while barred variables such as $\bar{y}$ and $\bar{M}$
denote reverse-mode cotangents, not complex conjugates.
The projected channels are
\begin{equation}
\begin{aligned}
H_R &= V_R^\dagger M V_R,
&
M V_R &\simeq V_R H_R,
\\
H_L &= V_L^\dagger M V_L,
&
M^\dagger V_L &\simeq V_L H_L^\dagger .
\end{aligned}
\label{eq:two-sided-krylov}
\end{equation}
Let $\theta_j^R$ and $\theta_i^L$ be the Ritz values of
$H_R$ and $H_L$, respectively. We define the scalar divided
difference by
\begin{equation}
F^{[1]}(x,y)
=
\begin{cases}
\dfrac{F(x)-F(y)}{x-y}, & x\ne y,\\[1.2ex]
F'(x), & x=y .
\end{cases}
\label{eq:scalar-divided-difference}
\end{equation}
The rectangular Daleckii--Krein table used in the two-sided
projection is then
\begin{equation}
F^{[1]}_{ij}
=
F^{[1]}(\theta_i^L,\theta_j^R).
\label{eq:daleckii}
\end{equation}
In the implementation, Ritz values that are indistinguishable
under a relative clustering tolerance of $10^{-10}$ are treated
by the derivative branch.

In the Ritz bases of $H_L$ and $H_R$, the projected
Fr\'echet map acts by entrywise multiplication with
$F^{[1]}_{ij}$. Its adjoint conjugates the divided-difference
table and the Ritz-coordinate transformations, producing the
small Arnoldi-coordinate cotangent $\bar{H}$. The cotangent
with respect to the full channel is then lifted as
\begin{equation}
\bar{M} \simeq V_L \bar{H} V_R^\dagger .
\label{eq:matrix-cotangent-lift}
\end{equation}

\bibliography{refs}

\end{document}